\documentstyle[twoside]{article}
\input{epsf.sty}

\catcode`\@=11
\long\def\@makefntext#1{
\protect\noindent \hbox to 3.2pt {\hskip-.9pt  
$^{{\eightrm\@thefnmark}}$\hfil}#1\hfill}		%CAN BE USED 

\def\thefootnote{\fnsymbol{footnote}}
\def\@makefnmark{\hbox to 0pt{$^{\@thefnmark}$\hss}}	%ORIGINAL 
	
\def\ps@myheadings{\let\@mkboth\@gobbletwo
\def\@oddhead{\hbox{}
\rightmark\hfil\eightrm\thepage}   
\def\@oddfoot{}\def\@evenhead{\eightrm\thepage\hfil
\leftmark\hbox{}}\def\@evenfoot{}
\def\sectionmark##1{}\def\subsectionmark##1{}}

\oddsidemargin=\evensidemargin
\renewcommand{\thefootnote}{\fnsymbol{footnote}}

\newcounter{sectionc}\newcounter{subsectionc}\newcounter{subsubsectionc}
\renewcommand{\section}[1] {\vspace{12pt}\addtocounter{sectionc}{1} 
\setcounter{subsectionc}{0}\setcounter{subsubsectionc}{0}\noindent 
	{\tenbf\thesectionc. #1}\par\vspace{5pt}}
\renewcommand{\subsection}[1] {\vspace{12pt}\addtocounter{subsectionc}{1} 
	\setcounter{subsubsectionc}{0}\noindent 
	{\bf\thesectionc.\thesubsectionc. {\kern1pt \bfit #1}}\par\vspace{5pt}}
\renewcommand{\subsubsection}[1] {\vspace{12pt}\addtocounter{subsubsectionc}{1}
	\noindent{\tenrm\thesectionc.\thesubsectionc.\thesubsubsectionc.
	{\kern1pt \tenit #1}}\par\vspace{5pt}}
\newcommand{\nonumsection}[1] {\vspace{12pt}\noindent{\tenbf #1}
	\par\vspace{5pt}}

\newcounter{appendixc}
\newcounter{subappendixc}[appendixc]
\newcounter{subsubappendixc}[subappendixc]
\renewcommand{\thesubappendixc}{\Alph{appendixc}.\arabic{subappendixc}}
\renewcommand{\thesubsubappendixc}
	{\Alph{appendixc}.\arabic{subappendixc}.\arabic{subsubappendixc}}

\renewcommand{\appendix}[1] {\vspace{12pt}
        \refstepcounter{appendixc}
        \setcounter{figure}{0}
        \setcounter{table}{0}
        \setcounter{lemma}{0}
        \setcounter{theorem}{0}
        \setcounter{corollary}{0}
        \setcounter{definition}{0}
        \setcounter{equation}{0}
        \renewcommand{\thefigure}{\Alph{appendixc}.\arabic{figure}}
        \renewcommand{\thetable}{\Alph{appendixc}.\arabic{table}}
        \renewcommand{\theappendixc}{\Alph{appendixc}}
        \renewcommand{\thelemma}{\Alph{appendixc}.\arabic{lemma}}
        \renewcommand{\thetheorem}{\Alph{appendixc}.\arabic{theorem}}
        \renewcommand{\thedefinition}{\Alph{appendixc}.\arabic{definition}}
        \renewcommand{\thecorollary}{\Alph{appendixc}.\arabic{corollary}}
        \renewcommand{\theequation}{\Alph{appendixc}.\arabic{equation}}
        \noindent{\tenbf Appendix \theappendixc #1}\par\vspace{5pt}}
\newcommand{\subappendix}[1] {\vspace{12pt}
        \refstepcounter{subappendixc}
        \noindent{\bf Appendix \thesubappendixc. {\kern1pt \bfit #1}}
	\par\vspace{5pt}}
\newcommand{\subsubappendix}[1] {\vspace{12pt}
        \refstepcounter{subsubappendixc}
        \noindent{\rm Appendix \thesubsubappendixc. {\kern1pt \tenit #1}}
	\par\vspace{5pt}}

\newcommand{\textlineskip}{\baselineskip=13pt}
\newcommand{\smalllineskip}{\baselineskip=10pt}

\def\eightcirc{
\begin{picture}(0,0)
\put(4.4,1.8){\circle{6.5}}
\end{picture}}
\def\eightcopyright{\eightcirc\kern2.7pt\hbox{\eightrm c}} 

\newcommand{\copyrightheading}[1]
	{\vspace*{-2.5cm}\smalllineskip{\flushleft
	{\footnotesize International Journal of Modern Physics C, #1}\\
	{\footnotesize $\eightcopyright$\, World Scientific Publishing
	 Company}\\
	 }}

\newcommand{\publisher}[2]{{\begin{center}\footnotesize\smalllineskip 
	Received #1\\
	Revised #2
	\end{center}
	}}

\def\abstracts#1#2#3{{
	\centering{\begin{minipage}{4.5in}\baselineskip=10pt\footnotesize
	\parindent=0pt #1\par 
	\parindent=15pt #2\par
	\parindent=15pt #3
	\end{minipage}}\par}} 

\newcommand{\bibit}{\nineit}
\newcommand{\bibbf}{\ninebf}
\renewenvironment{thebibliography}[1]
        {\frenchspacing
	 \ninerm\baselineskip=11pt
         \begin{list}{\arabic{enumi}.}
        {\usecounter{enumi}\setlength{\parsep}{0pt}     
	 \setlength{\leftmargin 12.7pt}{\rightmargin 0pt} %FOR 1--9 ITEMS
         \setlength{\itemsep}{0pt} \settowidth
	{\labelwidth}{#1.}\sloppy}}{\end{list}}

\newcounter{itemlistc}
\newcounter{romanlistc}
\newcounter{alphlistc}
\newcounter{arabiclistc}

\newcommand{\fcaption}[1]{
        \refstepcounter{figure}
        \setbox\@tempboxa = \hbox{\footnotesize Fig.~\thefigure. #1}
        \ifdim \wd\@tempboxa > 5in
           {\begin{center}
        \parbox{5in}{\footnotesize\smalllineskip Fig.~\thefigure. #1}
            \end{center}}
        \else
             {\begin{center}
             {\footnotesize Fig.~\thefigure. #1}
              \end{center}}
        \fi}

\newcommand{\tcaption}[1]{
        \refstepcounter{table}
        \setbox\@tempboxa = \hbox{\footnotesize Table~\thetable. #1}
        \ifdim \wd\@tempboxa > 5in
           {\begin{center}
        \parbox{5in}{\footnotesize\smalllineskip Table~\thetable. #1}
            \end{center}}
        \else
             {\begin{center}
             {\footnotesize Table~\thetable. #1}
              \end{center}}
        \fi}

\def\@citex[#1]#2{\if@filesw\immediate\write\@auxout
	{\string\citation{#2}}\fi
\def\@citea{}\@cite{\@for\@citeb:=#2\do
	{\@citea\def\@citea{,}\@ifundefined
	{b@\@citeb}{{\bf ?}\@warning
	{Citation `\@citeb' on page \thepage \space undefined}}
	{\csname b@\@citeb\endcsname}}}{#1}}

\newif\if@cghi
\def\cite{\@cghitrue\@ifnextchar [{\@tempswatrue
	\@citex}{\@tempswafalse\@citex[]}}
\def\citelow{\@cghifalse\@ifnextchar [{\@tempswatrue
	\@citex}{\@tempswafalse\@citex[]}}
\def\@cite#1#2{{$\null^{#1}$\if@tempswa\typeout
	{IJCGA warning: optional citation argument 
	ignored: `#2'} \fi}}

\def\pmb#1{\setbox0=\hbox{#1}
	\kern-.025em\copy0\kern-\wd0
	\kern.05em\copy0\kern-\wd0
	\kern-.025em\raise.0433em\box0}

\def\fnt#1#2{\footnotetext{\kern-.3em
	{$^{\mbox{\scriptsize #1}}$}{#2}}}

\def\fpage#1{\begingroup
\voffset=.3in
\thispagestyle{empty}\begin{table}[b]\centerline{\footnotesize #1}
	\end{table}\endgroup}

\def\runninghead#1#2{\pagestyle{myheadings}
\markboth{{\protect\footnotesize\it{\quad #1}}\hfill}
{\hfill{\protect\footnotesize\it{#2\quad}}}}
\font\tenrm=cmr10
\font\tenit=cmti10 
\font\tenbf=cmbx10
\font\bfit=cmbxti10 at 10pt
\font\ninerm=cmr9
\font\nineit=cmti9
\font\ninebf=cmbx9
\font\eightrm=cmr8

\def\qed{\hbox{${\vcenter{\vbox{			%HOLLOW SQUARE
   \hrule height 0.4pt\hbox{\vrule width 0.4pt height 6pt
   \kern5pt\vrule width 0.4pt}\hrule height 0.4pt}}}$}}

\renewcommand{\thefootnote}{\fnsymbol{footnote}}	%USE SYMBOLIC FOOTNOTE

\def\bsc{{\sc a\kern-6.4pt\sc a\kern-6.4pt\sc a}}  	%LATEX LOGO
\def\bflatex{\bf L\kern-.30em\raise.3ex\hbox{\bsc}\kern-.14em 
T\kern-.1667em\lower.7ex\hbox{E}\kern-.125em X} 

\begin{document}

\runninghead{Monte Carlo Simulations of Doped, Diluted Magnetic
Semiconductors} 
{Monte Carlo Simulations of Doped, Diluted Magnetic Semiconductors}

\normalsize\textlineskip
\thispagestyle{empty}
\setcounter{page}{1}

\copyrightheading{}			%{Vol. 0, No. 0 (1993) 000--000}

\vspace*{0.88truein}

\fpage{1}
\centerline{\bf Monte Carlo Simulations of Doped, Diluted Magnetic
Semiconductors}
\vspace*{0.035truein}
\centerline{\bf -- a System with Two Length Scales}
\vspace*{0.37truein}
\centerline{\footnotesize R. N. Bhatt and Xin Wan}
\vspace*{0.015truein}
\centerline{\footnotesize\it Department of Electrical Engineering, 
Princeton University}
\baselineskip=10pt
\centerline{\footnotesize\it Princeton, New Jersey 08544, U. S. A.}
%\vspace*{10pt}
%\centerline{\normalsize and}
%\vspace*{10pt}
%\centerline{\footnotesize SECOND AUTHOR}
%\vspace*{0.015truein}
%\centerline{\footnotesize\it Group, Laboratory, Address}
%\baselineskip=10pt
%\centerline{\footnotesize\it City, State ZIP/Zone, Country}
\vspace*{0.225truein}
\publisher{(received date)}{(revised date)}

\vspace*{0.21truein}
\abstracts{We describe a Monte Carlo simulation study of the magnetic
phase diagram of diluted magnetic semiconductors doped with shallow
impurities in the low concentration regime. 
We show that because of a wide distribution of interaction strengths,
the system exhibits strong quantum effects in the magnetically ordered
phase.
A discrete spin model, found to closely approximate the quantum system,
shows long relaxation times, and the need for specialized cluster
algorithms for updating spin configurations. 
Results for a representative system are presented.}{}{}

%\vspace*{10pt}
%\keywords{The contents of the keywords}

\textheight=7.8truein
\setcounter{footnote}{0}
\renewcommand{\thefootnote}{\alph{footnote}}

%\textlineskip			%) USE THIS MEASUREMENT WHEN THERE IS
%\vspace*{12pt}			%) NO SECTION HEADING

\vspace*{1pt}\textlineskip	%) USE THIS MEASUREMENT WHEN THERE IS
\section{Introduction}	%) A SECTION HEADING
\vspace*{-0.5pt}
\noindent
Diluted Magnetic Semiconductors (DMS) doped with a small concentration
of (shallow) charged impurities constitute an interesting magnetic system
which have a number of novel features for study by numerical simulation
techniques\cite{furdyna,averous}.  
These features arise because of the existence of two widely separated
length scales -- one of atomic dimensions ($\sim$~2~\AA) coming from the
$d$ electrons of the magnetic ion, and the other from the effective Bohr
radius of the shallow impurity, which is typically $\sim$~20~\AA.
In a certain concentration regime described below, these two lengths
conspire to produce a very wide distribution of energy scales in the
system, which poses special challenges for numerical simulations,
in particular the problem of long equilibration times.

A prototypical DMS system is a II-VI semiconductor such as CdTe or
ZnSe, which is in the zinc blende crystal structure, with some of
the divalent sites (Cd/Zn) substituted by a magnetic ion, most
commonly Mn. Manganese has two 4s electrons, and so is isovalent
with Cd or Zn, but it also has a half filled $3d$ shell, which by
Hund's rule, leads to a $S = 5/2$ ground state. These spins interact
via a short range exchange coupling in the semiconducting (insulating)
host, with nearest and second nearest neighbors usually most important. 
For concentrations $x$ of Mn in, say, Cd$_{1-x}$Mn$_x$Te, in excess of
around 0.2, the Mn spins are connected in a percolated network of short
range exchange couplings, and the system is found to undergo a
transition from a paramagnetic state to what is believed to be a spin
glass state, at a temperature which depends on $x$, but is of order 10 K. 
For $x$ well below that, the Mn spin system does not
percolate, and the magnetic behavior is well described in terms of
predominantly isolated Mn spins, and a statistically calculable number of
small clusters of Mn with typically 2, 3 or 4 Mn ions.

When shallow dopants are introduced in the system, they are characterized by
a single electron or hole bound in a hydrogenic $1s$ state, the electron/hole
binding energy scale being $\sim$~20~meV.
They interact with the Mn $d$-electron spin via an exchange interaction
of the Heisenberg type, which has a typical magnitude $\sim$~2~meV. Thus, it
is a reasonable approximation to neglect the modification of the dopant 
electron (or hole\footnote{To simplify the discussion we will consider
the electron case where the electronic spin is $s= 1/2$; the hole case
with spin-orbit coupling and $s= 3/2$ is more complex, but not essential
for the discussion here. 
Therefore, we will henceforth talk only about the $n$-doped case.} 
) wavefunction by the presence of Mn spins, and the low
energy Hamiltonian in the limit of low dopant density can be described
as a sum of pairwise exchange interactions between the electron spins 
(${\bf s}_i$) and the Mn spins (${\bf S}_j$)\cite{pwolff}:
\begin{equation}
\label{eqn:Hamiltonian}
{\cal H} = \sum_{i,j} J({\bf r}_i, {\bf R}_j) {\bf s}_i \cdot {\bf S}_j,
\end{equation}
where the sum is over Mn spins at ${\bf R}_j$, and electronic spins with 
wavefunctions centered around the donor sites at ${\bf r}_i$. 
The exchange interaction energy $J({\bf r}_i, {\bf R}_j)$ between the
electron centered at ${\bf r}_i$ and Mn spin at ${\bf R}_j$ is
proportional to the electronic charge density at the Mn site, i.e. 
proportional to $|\psi({\bf r}_i - {\bf R}_j)|^2$ , 
$\psi({\bf r})$ being the ground state wavefunction 
of the dopant electron. 
(This is very much like the contact interaction
between nuclear spins and electronic spins, since here the electron spin
has much greater extent than the spins formed by the Mn 3$d$ electrons.)
For shallow donors with hydrogenic $1s$ wavefunctions, therefore, 
\begin{equation}
\label{eqn:Exchange}
J({\bf r}_i, {\bf R}_j) = J_0 e^{- 2 |{\bf r}_i - {\bf R}_j| / a_B},
\end{equation}
where $J_0$ is the exchange constant and $a_B$ the Bohr radius 
($\sim$~20~\AA). 

For a single dopant, this results in a polarization of the Mn spins at
low temperatures in a ferromagnetic configuration with respect to
each other in the absence of Mn-Mn exchange. For dilute concentration
of Mn, so there are only a small fraction of near neighbor Mn clusters,
we may neglect the direct Mn-Mn interaction, this magnetic polaron around
each donor electron survives, and in previous work one of us has 
shown\cite{pwolff} that the relative orientation of
the polarons at still lower temperatures becomes ferromagnetic, as
the radius of the polaron increases with lowering of $T$.
The system would be expected to show a genuine ferromagnetic
transition when the magnetic polarons percolate, so that the ferromagnetic
order becomes genuinely long ranged.

However, since the percolation fraction for randomly placed spheres in
three dimensions is $\sim$~20\%, even below the transition, many Mn spins
remain essentially unattached to the percolated cluster till much lower
temperatures. This results in a very unusual ferromagnetic phase, in
which a substantial part of the spin entropy survives down to
very low temperatures. Since analytic results are available only in
certain limiting cases\cite{bhatt}, we have
carried out Monte Carlo simulations of the doped DMS system for
a typical dopant concentration where the isolated donor approximation
for the ground state donor wavefunction would be reasonable, and
a concentration of Mn where the assumption of no direct Mn-Mn interaction
would be reasonable.

\section{Heisenberg Model}
\noindent
In most magnetic models involving Heisenberg spins with large quantum
numbers ($S = 5/2$ here for the Mn spins which dominate the magnetic
response), quantum corrections are small and the classical vector spin
model has been found to be adequate. 
Therefore, we performed Monte Carlo simulations on the model represented
by Eqs.~(\ref{eqn:Hamiltonian}) and (\ref{eqn:Exchange}) using classical
vector spins.

On a zinc-blende lattice with lattice constant of 5\AA,
dopants and magnetic ions are distributed randomly on anionic and
cationic sites of the lattice respectively.
We choose $a_B = 20 \AA$, $n_d = 10^{18} cm^{-3}$, and $x = 0.001$.
The interaction between electron spins and Mn spins is antiferromagnetic. 
The interaction between electron spins has been shown to be
unimportant\cite{pwolff}, while Mn-Mn couplings are not considered
because of the rare clusters of two Mn spins on the neighboring sites
($\sim$ 1\%) form a magnetically inert singlet at low temperatures. 
For these concentrations, Mn spins are coupled most strongly to only a
few electrons, so for each Mn spin we have neglected interactions which
are smaller than $10^{-5}$ times its strongest interaction. 
With this cutoff almost all Mn spins are coupled to at least two
electron spins. 

\begin{figure}[htbp]
\vspace*{-0.5cm}
\epsfxsize=5.0in
\begin{center}
\leavevmode
\epsffile{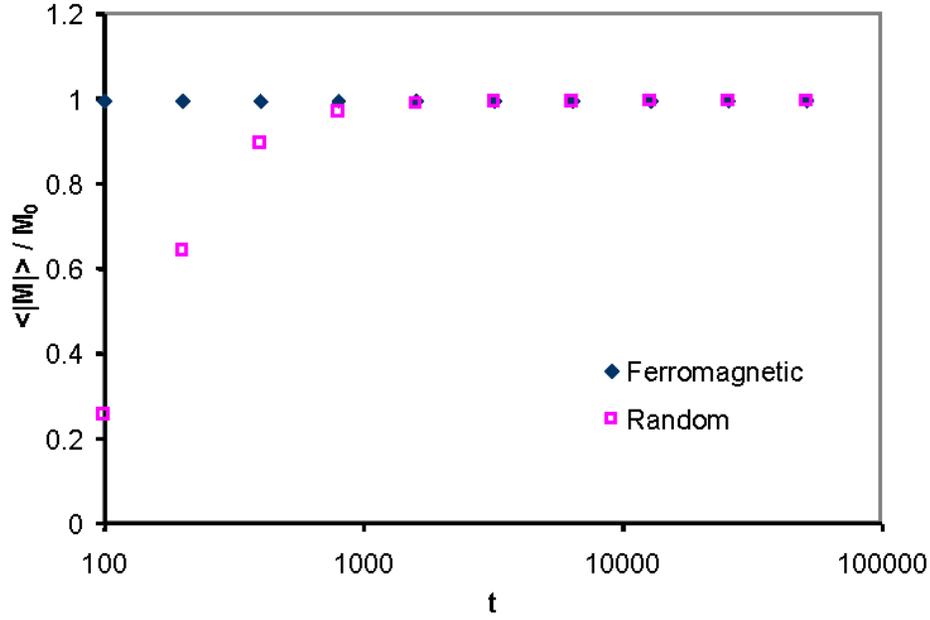}
\end{center}
\vspace*{-0.5cm}
\caption{
Sample averaged magnetization per spin $M / M_0$ as a function of time $t$
for a system of 864 Mn spins at $T = 0.00001$. 
The upper curve evolves from ferromagnetic initial spin configuration,
while the lower one evolves from random spin configuration. 
Classical vector spins are used in the simulation. 
\label{fig:HeisenbergT0.00001N27Equ}
}
\end{figure}

We carried out simulations on systems of linear sizes 20, 30, and 40,
which have 256, 864, and 2048 Mn spins and 8, 27, and 64 electron spins,
respectively\cite{wan}.  
Since we are simulating a very inhomogeneous system with large number of
spins, the criterion of equilibration is crucial, especially when we try 
to apply the simulation to higher Mn concentrations, when clusters of
many nearest neighboring Mn spins exist.
Therefore, we adopt a widely used scheme in simulating spin glasses to
determine the equilibration time\cite{Young}. 
Consider two replicas with identical locations of particles and 
identical couplings between them. 
We start one from a totally random spin configuration, and the other from a
ferromagnetic configuration, in which Mn spins point in
the same direction, while electron spins point opposite. 
We use the sample averaged magnitude of magnetization per Mn spin 
$\langle |M| \rangle$ as the quantity to converge.  
One expects that $\langle |M| \rangle$ increases from zero in the initially 
random replica, whereas it will decrease from the saturation value $M_0 = 5/2$ 
in the initially ferromagnetic replica.  
After $\langle |M| \rangle$ of the two replicas agree within error bars, 
one expects both systems have reached equilibrium, which determines the
time needed before measurements can be taken.

Figure~\ref{fig:HeisenbergT0.00001N27Equ} shows the equilibration of
a system of 864 Mn spins at $T = 0.00001$. 
The replica method turns out to be suitable to determine 
the equilibration time for the Heisenberg model even 
at a temperature as low as $0.00001$. 
In the simulation, we use the standard Metropolis algorithm in which
consecutive configurations can be different for no more than one single
spin. 
At each Monte Carlo step, we go through all the spins successively and
flip the spins with equilibrium probabilities.

\begin{figure}[htbp]
\vspace*{-0.5cm}
\epsfxsize=5.0in
\begin{center}
\leavevmode
\epsffile{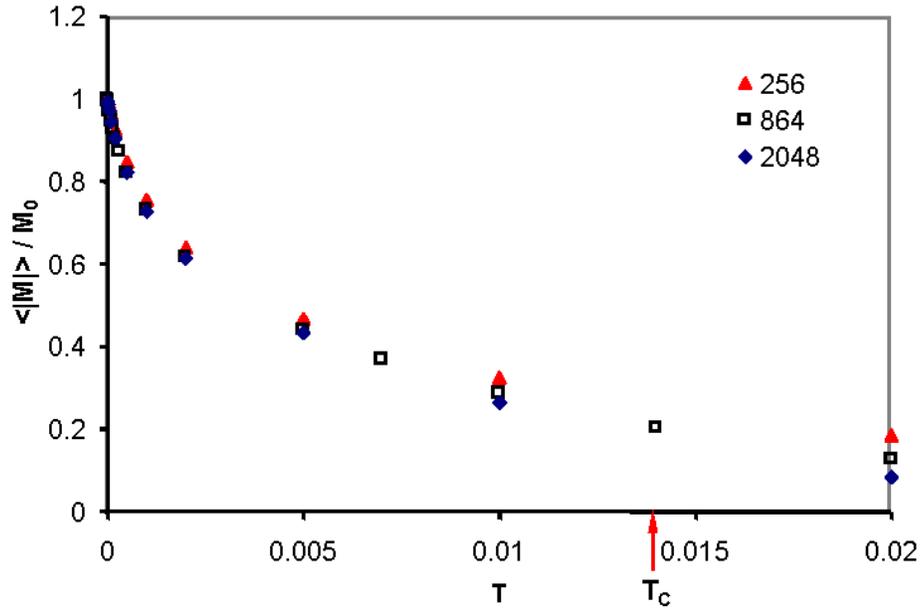}
\end{center}
\vspace*{-0.5cm}
\caption{
Normalized magnitude of magnetization per Mn spin $\langle |M| \rangle$ 
at different temperatures for systems of 256, 864, and 2048 Mn spins in
the Heisenberg model. 
\label{fig:HeisenbergM}
}
\end{figure}

Figure~\ref{fig:HeisenbergM} shows the sample averaged magnitude of 
magnetization per Mn spin as a function of temperature for systems 
of 256 to 2048 Mn spins. 
$\langle |M| \rangle$ looks very different from that of a conventional 
ferromagnet. 
At the lowest temperatures simulated for the classical Heisenberg model, 
$\langle |M| \rangle$ rises steeply, 
approaching $M_0 = 5/2$ in a linear fashion. 
Magnetization per Mn spin is a good estimate of the percentage of
Mn spins belonging to the percolating ferromagnetic cluster. 
The critical temperature $T_C \simeq 0.014$ is obtained by using various
cumulants of the magnetization, which will be discussed later in
Section 3 %\ref{discreteModel} 
for the discrete spin model.  

\begin{figure}[htbp]
\vspace*{-0.5cm}
\epsfxsize=5.0in
\begin{center}
\leavevmode
\epsffile{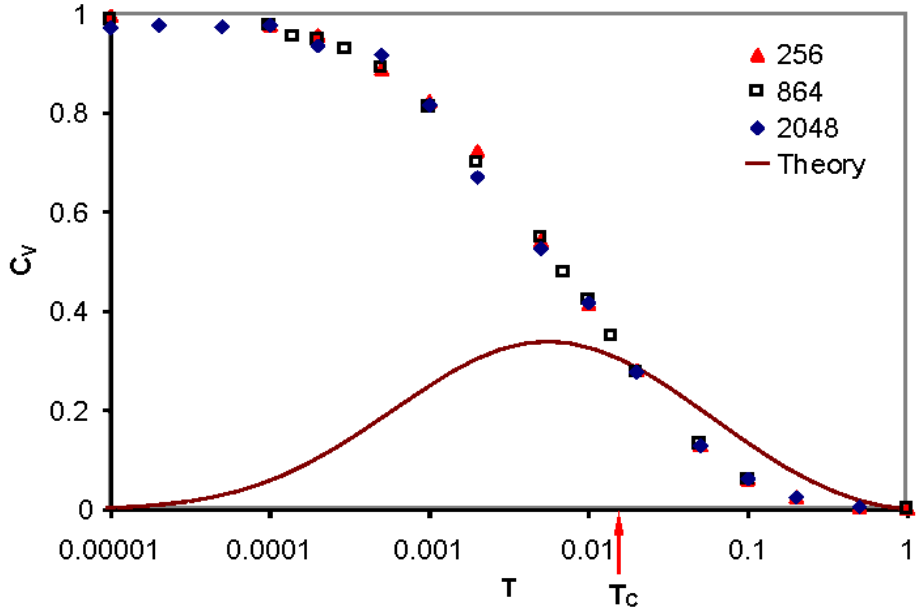}
\end{center}
\vspace*{-0.5cm}
\caption{
Specific heat $C_v$ as a function of temperature for systems of 256,
864, and 2048 Mn spins in the Heisenberg model.
The solid curve shows the theoretical estimate.  
\label{fig:HeisenbergCv}
}
\end{figure}

The specific heat per spin is plotted against temperature in
Fig.~\ref{fig:HeisenbergCv}.
At low temperatures, $C_v$ approaches unity which is the result of
the equipartition theorem. 
This suggests that we cannot treat Mn spins as classical Heisenberg 
objects. 
The solid curve in Fig.~\ref{fig:HeisenbergCv} is a theoretical
estimate of $C_v$ for the quantum case\cite{bhatt}, which is expected to hold 
at low temperatures (only), and this clearly deviates from the simulation 
results below $T_c$ and approaches zero in low temperature limit.  
It is noteworthy that this strong deviation between classical and
quantum spins already occurs at temperatures where the magnetization is
less than 40\% of the saturation value. 
Consequently quantum effects are important in this system. 

\newpage
\section{Discrete Spin Model}
\label{discreteModel}

\subsection{Discrete 12-state model}
\noindent
In order to simulate the effects of the discreteness of the quantum
spins in doped diluted magnetic semiconductors without having to perform
a fully quantum simulation, we have modified Mn spins from a continuous
vector spin to various discrete spin versions.
Here, we discuss the results of a discrete 12-state model, 
in which each spin can take one of the 12 $\langle 110 \rangle$
directions of a cubic lattice with a Hamiltonian given by
Eqs.~(\ref{eqn:Hamiltonian}) and (\ref{eqn:Exchange}).

\begin{figure}[htbp]
\vspace*{-0.5cm}
\epsfxsize=4.5in
\begin{center}
\leavevmode
\epsffile{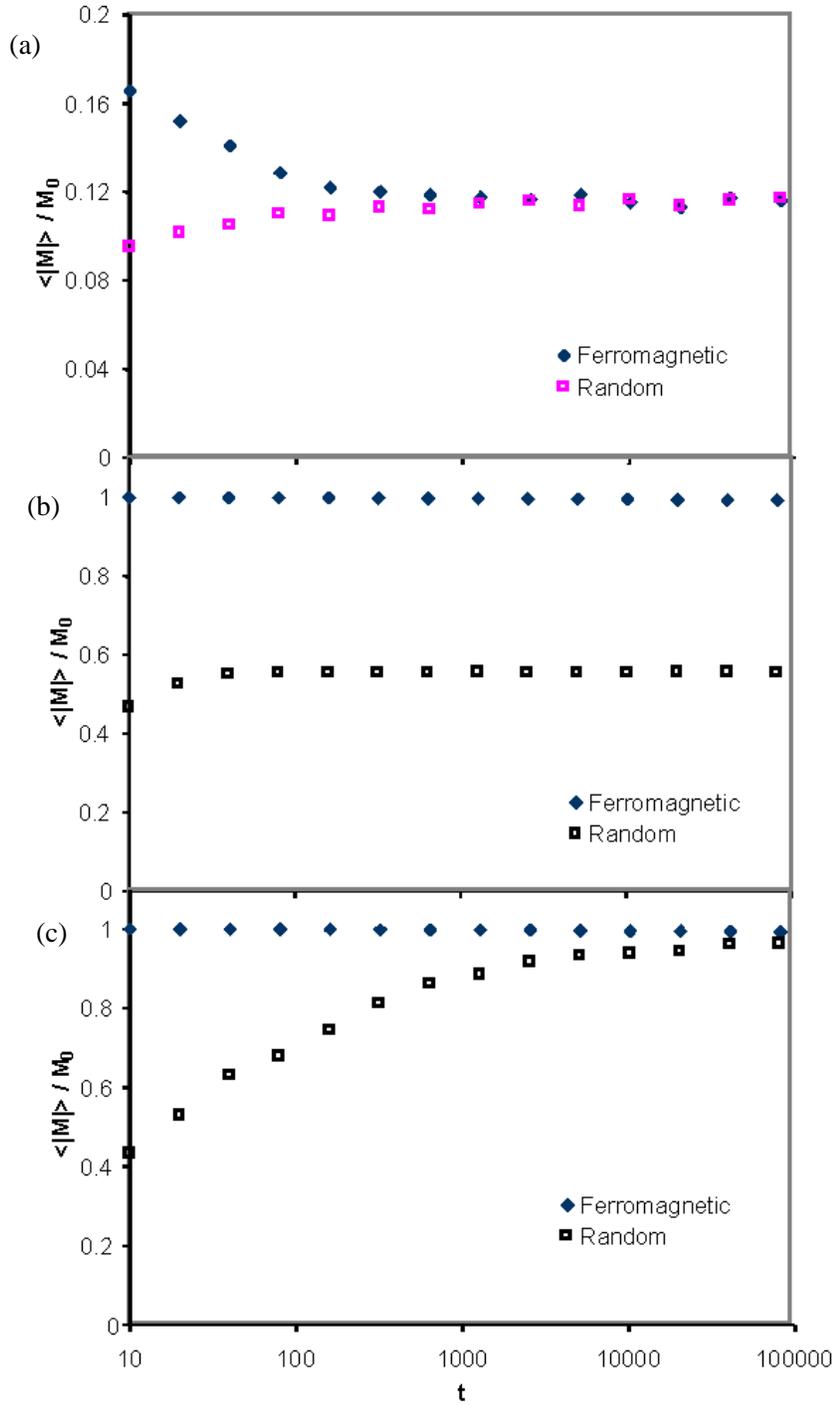}
\end{center}
\vspace*{-0.5cm}
\caption{
Magnetization per spin $M$ as a function of time $t$ for system with 256
Mn spins in replicas starting from ferromagnetic configuration and
random configuration in the discrete 12-spin model
(a) at $T = 0.04$ with single spin flips, 
(b) at $T = 0.00001$ with single spin flips, and 
(c) at $T = 0.00001$ with the cluster updating algorithm described in
the text.
\label{fig:Vertex12N8Equ}
}
\end{figure}

Figure~\ref{fig:Vertex12N8Equ}(a) and (b) compare the evolution of 
magnetization per Mn spin in  a system of 256 Mn spins at $T = 0.04$ and
$T = 0.00001$ using standard single spin flip dynamics. 
At the higher temperature, $\langle |M| \rangle$ of the two replicas
converge to the same value, implying the equilibrium is reached in both
replicas.  
However, at temperature three orders below, the magnetization values hardly
change after initial 100 steps and show no indication of equilibration. 
This is easily understood as follows:
each single spin flip is associated with a finite energy change. 
Therefore, the discretization of the energy levels will inevitably lead to 
significant slowing down of the simulation at low temperatures. 
The broad distribution of interactions worsen the equilibration of
discrete spins by forming domains around each electron spins where the
exchanges are strong. 
These domains are unlikely to flip in reasonable times, because the
exchanges at the domain surfaces are much weaker compared to that in the
bulk.  
A new algorithm, which can effectively flip each domains at a
single step, is thus required to achieve equilibration.

\subsection{Cluster algorithm}
\noindent
To develop a cluster algorithm for various spin models, we follow
Wolff's cluster algorithm\cite{uwolff}. 
We first define Ising-like spin-flip operation 
${\bf s} \rightarrow {\bf s'}$ as the reflection with respect to the
plane orthogonal to a unit vector $\hat{n}$,
\begin{equation}
{\cal R} (\hat{n}) {\bf s} = {\bf s} - 2 ({\bf s} \cdot \hat{n}) \hat{n},
\end{equation}
which satisfies
\begin{equation}
\label{idempotent}
{\cal R} (\hat{n})^2 =1,
\end{equation}
and
\begin{equation}
\label{invariant}
[ {\cal R} (\hat{n}) {\bf s}_1 ] \cdot [ {\cal R} (\hat{n}) {\bf s}_2 ]
	= {\bf s}_1 \cdot {\bf s}_2.
\end{equation}
In the discrete spin model, only finite number of unit vectors $\hat{n}$ exist 
that transform an arbitrary discrete spin into other discrete spins.
For instance, unit vectors along $\langle 110 \rangle$ and 
$\langle 100 \rangle$ directions are a set of appropriate reflections 
for discrete spins along $\langle 110 \rangle$ directions.

Due to the non-nearest-neighboring nature of spin exchanges in doped
magnetic semiconductors, the cluster algorithm needs extra consideration
whether the whole cluster is flipped or not. 
The revised cluster algorithm now consists of the following 
sequence of operations. 
(a) Choose a electron spin ${\bf s}_0$ at ${\bf r}_0$ and 
an appropriate reflection $\hat{n}$ randomly. 
${\bf s}_0$ is the seed of the cluster ${\cal C}$ to be built.
(b) Visit all manganese spins and activate the bond between 
the electron spin and a manganese spin ${\bf S}$ at ${\bf R}$ 
with probability
\begin{equation}
{\cal P}_{I} ({\bf S}) = 1 - {\rm exp} \{ {\rm min} [0, 
	2 J({\bf r}_0, {\bf R}) ({\bf s}_0 \cdot \hat{n}) 
	({\bf S} \cdot \hat{n}) / T] \}.
\end{equation}
If the bond is activated, add ${\bf S}$ to the cluster ${\cal C}$.
(c) Visit all electron spins, except the seed ${\bf s}_0$, and add an
electron spin ${\bf s}$ to the cluster ${\cal C}$ with probability
\begin{equation}
{\cal P}_{II} ({\bf s}) = 1 - {\rm exp} \{ {\rm min} [0, 
	\sum_{{\bf S} \in {\cal C}} 2 J({\bf r}, {\bf R}) ({\bf s} \cdot \hat{n}) 
	({\bf S} \cdot \hat{n}) / T] \}.
\end{equation}
(d) Flip the whole cluster ${\cal C}$ with probability
\begin{equation}
{\cal P}_{III} ({\cal C}) = {\rm exp} \{ {\rm min} [0, 
	\sum_{{\bf s} \in {\cal C} - \{{\bf s}_0\}, 
	{\bf S} \bar{\in} {\cal C}} 2 J({\bf r}, {\bf R})
	({\bf s} \cdot \hat{n}) ({\bf S} \cdot \hat{n}) / T] \}.
\end{equation}
Detailed balance is fulfilled since the transition probabilities ${\cal W}$ 
between two configurations $\{ {\bf s}, {\bf S} \}$ and 
$\{ {\bf s'}, {\bf S'} \}$ that differ by a flip ${\cal R} (\hat{n})$ 
on a cluster ${\cal C}$ built around electron spin ${\bf s}_0$ obey 
\begin{eqnarray}
& & \frac{{\cal W} (\{ {\bf s}, {\bf S} \} \rightarrow 
	\{ {\bf s'}, {\bf S'} \})}{{\cal W} (\{ {\bf s'}, {\bf S'} \} 
	\rightarrow \{ {\bf s}, {\bf S} \}) }
 = \frac{{\cal P}_{III} ({\cal C})
	\prod_{{\bf s} \bar{\in} {\cal C}} [1 - {\cal P}_{II} ({\bf s})] 
	\prod_{{\bf S} \bar{\in} {\cal C}} [1 - {\cal P}_{I} ({\bf S})]}
	{{\cal P}_{III} ({\cal C'})
	\prod_{{\bf s'} \bar{\in} {\cal C'}} [1 - {\cal P}_{II} ({\bf s'})] 
	\prod_{{\bf S'} \bar{\in} {\cal C'}} [1 - {\cal P}_{I} ({\bf S'})]}
	\nonumber \\
 &=& {\rm exp} \left [ {2 \over T} \left ( 
	\sum_{{\bf s} \in {\cal C} - \{{\bf s}_0\}, {\bf S} \bar{\in} {\cal C}} 
	+ \sum_{ {\bf s} \bar{\in} {\cal C}, {\bf S} \in {\cal C} }  
	+ \sum_{{\bf s} = {\bf s}_0, {\bf S} \bar{\in} {\cal C}} \right )
	J({\bf r}, {\bf R}) ({\bf s} \cdot \hat{n}) ({\bf S} \cdot \hat{n}) 
	\right ] \nonumber \\
 &=& {\rm exp} \left \{ - {1 \over T} \sum_{{\bf s}, {\bf S}} 
	J({\bf r}, {\bf R}) ({\bf s'} \cdot {\bf S'} - {\bf s} \cdot {\bf S})
	\right \}
\end{eqnarray}
It is worth pointing out that all probabilities for activating bonds 
within ${\cal C}$ are the same before or after the cluster flipping 
because of Eqs. (\ref{idempotent}) and (\ref{invariant}). 
Due to the extremely broad distribution of spin exchanges, we add 
the following step to reduce the time in which all the manganese spins 
far away from any electron spin can reach thermal equilibrium.   
(e) Visit all manganese spins and flip each individual spin ${\bf S}$ 
to an arbitrary discrete spin ${\bf S'}$ with probability
\begin{equation}
{\cal P}({\bf S}) = {\rm exp} \{ {\rm min} [0, \sum_{\bf s} 
	J({\bf r}, {\bf R}) {\bf s} \cdot ({\bf S} - {\bf S'})
	 / T] \}
\end{equation}
Ergodicity of the above processes is guaranteed by the fact that 
there is always nonzero probability that ${\cal C}$ consists of 
only one spin, and that there is always one reflection or 
two consecutive reflections connecting any two discrete spins.
This allows exploration of all configurations with finite probabilities
required for ergodicity.

The efficiency of the cluster algrithm can be tested at very low
temperatures. 
Figure~\ref{fig:Vertex12N8Equ}(c) shows the average evolution curve of
the systems of 256 Mn spins at $T = 0.00001$.
Cluster algorithm is applied in the random replica.
In 100,000 steps, $\langle |M| \rangle$ in the random replica rises to 
almost 97\% of the value in the ferromagnetic replica. 
The small discrepancy largely comes from the worst 10\% of the samples 
in which a longer time is required to remove all the domain walls.

The development of ordered state from initially random configuration 
particularly in systems of 256 Mn spins using cluster algorithm confirms that 
the replica with such ferromagnetic initial configuration can reach
equilibrium by single spin-flip dynamics. 
Obviously, the reason is that, for the simple Hamiltonian in 
Eq.~(\ref{eqn:Hamiltonian}), we know that the ground state is a
ferromagnetic state at zero temperature, since there is no frustration.
Therefore, even if the random replica is not in equilibrium, we can
measure the equilibrium quantities in the replica starting from the
appropriate initial configuration after enough many steps using only
single spin-flip algorithm. 

Although the cluster algorithm with the replica method 
determining the equilibration time may appear to be
a luxury in the dilute concentration limit, it is absolutely
necessary when sizable clusters with Mn-Mn interactions exist at higher
Mn concentrations when the ground state configuration is unknown.

\subsection{Results of the discrete 12-state model}
\noindent
In this subsection, we show results based on the assumption that the
replica starting from ferromagnetic initial configuration reaches
equilibrium after long enough time, using single spin-flip dynamics. 

\begin{figure}[htbp]
\vspace*{-0.5cm}
\epsfxsize=5.0in
\begin{center}
\leavevmode
\epsffile{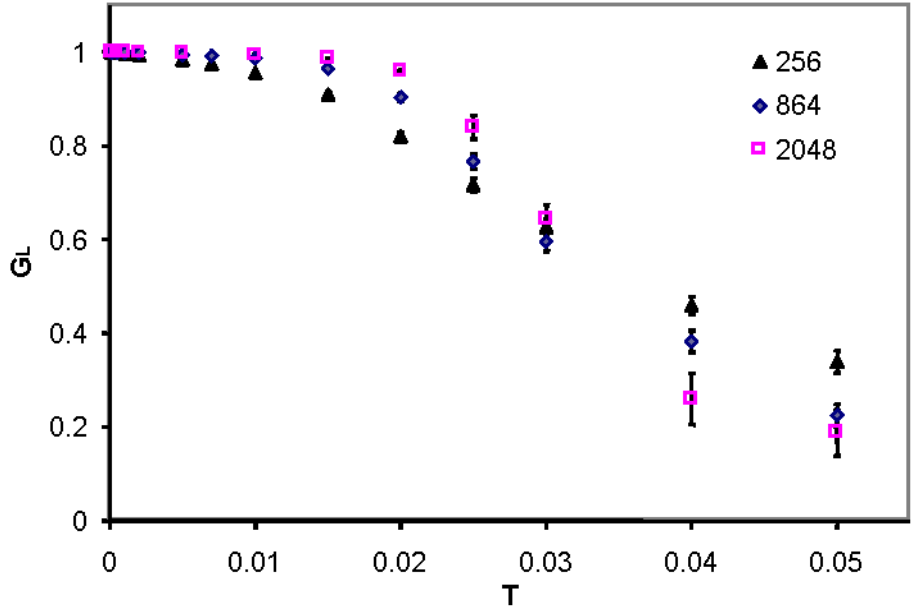}
\end{center}
\vspace*{-0.5cm}
\caption{
$G_L$ at different temperatures for systems of 256, 864, and 2048 Mn spins 
in the discrete 12-state model.  
The crossings of the three curves indicates $T_c \sim 0.03$.
\label{fig:Vertex12GL}
}
\end{figure}

A dimensionless quantity\cite{binder} used to characterize the
ferromagnetic order is given by 
\begin{equation}
\label{cumulant}
G_L = {1 \over 2} \left \{ 5 - 3 { \langle M^4 \rangle \over 
	\langle M^2 \rangle^2 } \right \}.
\end{equation}
The coefficients in Eq.~(\ref{cumulant}) are chosen so that $G_L$
approaches zero in the thermodynamic limit above $T_c$ and unity below $T_c$. 
From finite size scaling theory, we expect that $G_L$ is size
independent at the critical temperature. 
The three $G_L$ curves in Fig.~\ref{fig:Vertex12GL} seem to cross at $T_c
\simeq 0.03$. 
Below $T_c$ $G_L$ increases as system size increases, while $G_L$
decreases as system size increases above $T_c$, as expected.   

\begin{figure}[htbp]
\vspace*{-0.5cm}
\epsfxsize=5.0in
\begin{center}
\leavevmode
\epsffile{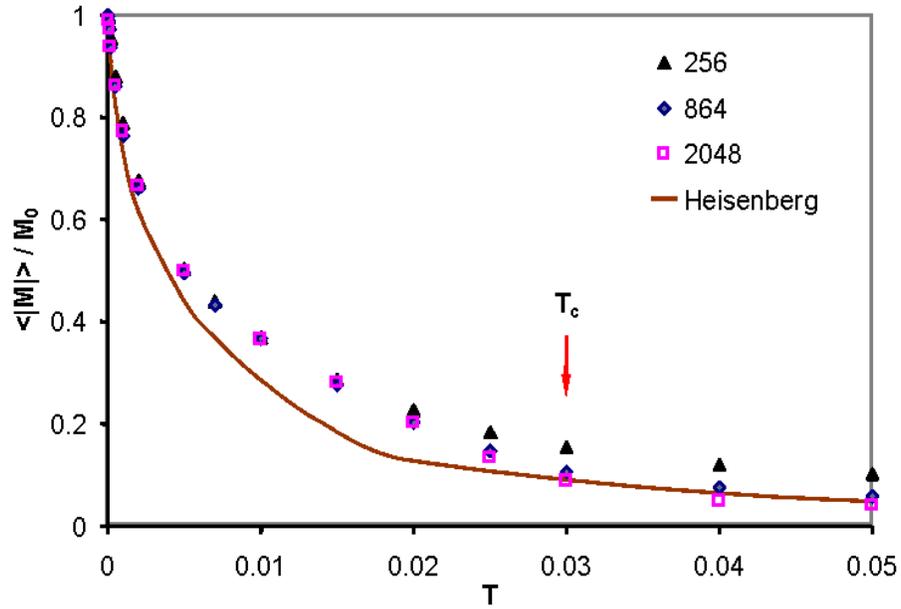}
\end{center}
\vspace*{-0.5cm}
\caption{
Normalized magnitude of magnetization per Mn spin versus temperature for
different system sizes in the discrete 12-state model.
The solid line fits $\langle |M| \rangle$ of the system of 864
Mn spins in the Heisenberg model. 
\label{fig:Vertex12M}
}
\end{figure}

\begin{figure}[htbp]
\vspace*{-0.5cm}
\epsfxsize=5.0in
\begin{center}
\leavevmode
\epsffile{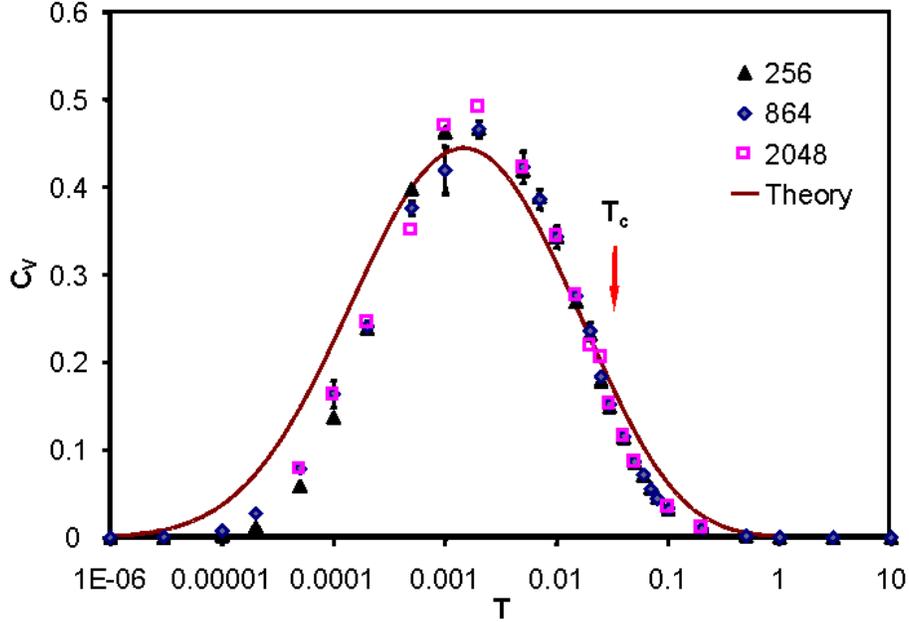}
\end{center}
\vspace*{-0.5cm}
\caption{
Specific heat $C_v$ as a function of temperature for different system
sizes in the discrete 12-state model. 
The solid curve is again the theoretical estimate of $C_v$.
\label{fig:Vertex12Cv}
}
\end{figure}

The magnetization per Mn spin is plotted versus temperature for the
three different system sizes in Fig~\ref{fig:Vertex12M}. 
Compared with the results in the Heisenberg model, the curves are
roughly scaled by a factor of 2 on the temperature axis, as is $T_c$. 
The specific heat per spin shown in Fig.~\ref{fig:Vertex12Cv} starts
dropping at $T \simeq 0.001$.
The simulation results in the discrete 12-state model agree quite well
with the theoretical estimate for the specific heat of the quantum
system\cite{bhatt}.    

\newpage
\section{Conclusion}
\noindent
In conclusion, we have presented results of a numerical
simulation appropriate for diluted magnetic semiconductors with
a low concentration of magnetic ions ($x$) and a low concentration of
shallow dopants ($n_d$).
The restriction on $n_d$ is provided by the requirement that the system
remain insulating, i.e. below the Mott concentration 
(or explicitly $n_d \cdot a_B^3 < 0.016$).
The restriction on $x$ is required for the Mn-dopant interaction to
be the dominant interaction. 

In this regime, the system is found to order ferromagnetically at
low temperatures, but the ferromagnetic state is rather unusual, in
that the magnetization does not approach the saturation value till well
below the ordering temperature. 
This arises because of the effective couplings that characterize the
system are distributed
on a logarithmic scale owing to the exponential variation of the
interactions with distance. 
Because of this wide distribution, the
system cannot be adequately represented by a classical Heisenberg model
at the low temperatures of interest. 
A discrete spin model is found
to remove most of the inadequacies of the continuous spin model,
but requires the application of specialized cluster algorithms
to help alleviate the extremely long relaxation times that are inherent
in a discrete spin model.

While there may be other methods to get around the problems of
equilibration in the ferromagnetic regime, the cluster algorithms
described above, which are found to work well (at least for small sizes)
in the concentration regimes studied, should in principle be useful in
other regimes, e.g. higher Mn concentration, where Mn-Mn interactions
cannot be neglected. 
The higher concentration regimes will involve frustration effects
as well, and having a reliable numerical method to ensure equilibration
is prerequisite for any study of such effects in the diluted
magnetic semiconductors.
We are currently testing if big enough sizes can be brought to
equilibrium with the above scheme to enable finite size scaling for
accurate location of $T_c$, and also to ensure that finite size
corrections for the susceptibility, specific heat and other
thermodynamic quantities in the ferromagnetic phase remain small.  

\nonumsection{Acknowledgement}
\noindent
This work was supported by NSF DMR 9809483.

\nonumsection{References}
%\noindent

%\appendix
%\noindent


\begin{thebibliography}{000}
\bibitem{furdyna}
P. A. Wolff, in {\bibit Semiconductors and Semimetals},
ed. J. K. Furdyna and J. Kossut (Academic, San Diago, 1988), Vol. 25,
p. 413.

\bibitem{averous}
P. A. Wolff, in {\bibit Semimagnetic and Diluted Magnetic Semiconductors},
ed. M. Averous and B. Balkanski (Plenum, New York, 1991).

\bibitem{pwolff}
P.A. Wolff, R. N. Bhatt, and A. C. Durst, 
{\bibit J. Appl. Phys.} {\bibbf 79}, 5196 (1996).

\bibitem{bhatt}
R. N. Bhatt, {\bibit unpublished}.

\bibitem{Young}
R. N. Bhatt and A. P. Young, 
{\bibit Phys. Rev.} {\bibbf B 37}, 5606 (1988).

\bibitem{wan}
For more details, see Xin Wan and R. N. Bhatt, {\bibit unpublished}.

\bibitem{uwolff}
U. Wolff, {\bibit Phys. Rev. Lett.}, {\bibbf 62}, 361 (1989).

\bibitem{binder}
K. Binder and A. P. Young, 
{\bibit Rev. Mod. Phys.} {\bibbf 58}, 801 (1986).

\end{thebibliography}
\end{document}